\documentclass[aps,prd,eprint,showpacs,superscriptaddress,showkeys,floatfix,nofootinbib,11pt]{revtex4-1}
\pdfoutput=1
\usepackage{graphicx}% Include figure files
\usepackage{dcolumn}% Align table columns on decimal point
\usepackage{bm}% bold math
\usepackage{xcolor}% bold math
\usepackage{url}
\usepackage{amsmath,amssymb,graphicx}
\usepackage{amsmath}
\usepackage{amssymb}
\usepackage{mathrsfs}
\usepackage{accents}
\usepackage{epsfig}
\usepackage{slashed}
\usepackage[lofdepth,lotdepth,caption=false]{subfig}
\usepackage{mathrsfs}
\usepackage{feynmp}
\usepackage[lofdepth,lotdepth,caption=false]{subfig}
\usepackage{soul}
\usepackage{hyperref}
\usepackage[normalem]{ulem}

\newlength\figureheight 
\newlength\figurewidth 
\newcommand{\inv}[1]{{#1}^{-1}}

\newcommand{\ds}[1]{{\displaystyle #1}}

\begin{document}

\title{A Neutrinophilic 2HDM as a UV Completion for the Inverse Seesaw Mechanism}
\author{Enrico Bertuzzo}
\affiliation{Departamento de F\'{\i}sica Matem\'atica, Instituto de F\'{\i}sica,
 Universidade de S\~ao Paulo, C.\ P.\ 66.318, 05315-970 S\~ao Paulo, Brazil}
\author{Pedro A. N. Machado}
\affiliation{Theoretical Physics Department, Fermi National Accelerator Laboratory, Batavia, IL, 60510, USA} 
 \author{Zahra  Tabrizi}
\affiliation{Departamento de F\'{\i}sica Matem\'atica, Instituto de F\'{\i}sica,
 Universidade de S\~ao Paulo, C.\ P.\ 66.318, 05315-970 S\~ao Paulo, Brazil}
\author{Renata Zukanovich Funchal}
\affiliation{Departamento de F\'{\i}sica Matem\'atica, Instituto de F\'{\i}sica,
 Universidade de S\~ao Paulo, C.\ P.\ 66.318, 05315-970 S\~ao Paulo, Brazil}

\date{\today} 

%%%%%%%%%%%%%%%%%%%%%%%%%%%%%%%%%%%%%%%%%%%%%%%%%%%%%%%%%%%%


\begin{abstract}

In Neutrinophilic Two Higgs Doublet Models, Dirac neutrino masses are
obtained by forbidding a Majorana mass term for the right-handed
neutrinos via a symmetry.  We study a variation of such models in
which that symmetry is taken to be a local $U(1)$, leading naturally
to the typical Lagrangian of the inverse seesaw scenario.  The
presence of a new gauge boson and of an extended scalar sector result in
a rich phenomenology, including modifications to $Z$, Higgs 
and kaon decays as well as to electroweak precision parameters, and a
pseudoscalar associated to the breaking of lepton number.

\end{abstract}

\begin{flushright}
	FERMILAB-PUB-17-210-T
\end{flushright}

\maketitle

%%%%%%%%%%%        I N T R O D U C T I O N        %%%%%%%%%%
%%%%%%%%%%%%%%%%%%%%%%%%%%%%%%%%%%%%%%%%
%%%%%%%%%%%%%%%%%%%%%%%%%%%%%%%%%%%%%%%%
\section{Introduction}

More than two decades of neutrino experiments have confirmed that
neutrinos are massive particles and oscillate. Strong bounds from
cosmological~\cite{Ade:2015xua} and terrestrial~\cite{Aseev:2011dq}
experiments suggest that neutrino masses should be below the eV
scale. This may be an indication that the mechanism behind the
generation of neutrino masses may be different from the Higgs
mechanism of the Standard Model (SM).

The seesaw mechanism is one of the most elegant and economic ways to
explain the smallness of neutrino
masses~\cite{Minkowski:1977sc,Mohapatra:1979ia,Schechter:1980gr}. In
the Type-I seesaw, the SM particle content is enlarged by at least two
heavy right-handed neutrino singlets, $N_i$ ($i$ denotes the
generation), such that the neutrino mass generation mechanism reads
\begin{equation}
	-{\cal L}_\nu = y^\nu_{ij} \; \overline{\ell}_{iL} \tilde{H} N_j + \frac{1}{2} M_R^{i} \overline{N^c_i} N_i\, + {\rm h.c.}\, ,
\end{equation}
where $\ell_{iL}$ are the SM lepton doublets, $H$ is the SM Higgs
scalar doublet and the tilde denotes charge conjugation. The heavy
neutrino mass scale suppresses the Dirac mass term contribution
typically resulting in active neutrino masses $m_\nu \simeq - v^2
y^{\nu\, T} \inv{M_R} y^\nu$, where $v=246$~GeV is the electroweak
(EW) scale. To avoid tiny Yukawa couplings (of order $10^{-11}$ or so)
and still have neutrino masses in the sub-eV range, the right-handed
neutrinos should be at a very high scale, ${\cal O}(10^{13})$~GeV,
which makes the model generally inaccessible to experiments except perhaps to 
future neutrinoless double beta decay measurements. This remains true
even if the right-handed neutrinos are taken to be at the TeV scale,
with $\mathcal{O}(10^{-6})$ Yukawas, since the light-heavy neutrino
mixing would be of the order of the ratio of light to heavy neutrino
masses, that is, $\theta\sim\mathcal{O}(10^{-12})$.
 
 An interesting variation of this scheme, which may yield observable
 phenomenology at colliders, is provided by the (double) inverse seesaw
 mechanism~\cite{Wyler:1982dd,Mohapatra:1986bd,GonzalezGarcia:1988rw}. In
 this scenario, in addition to the three right-handed neutrinos, three
 left-handed SM singlets, $\psi_{L}^{i}$, are introduced, with a
 Lagrangian
\begin{equation}\label{eq:inv_seesaw}
	-{\cal L}_\nu = y^\nu_{ij} \; \overline{\ell}_{iL} \tilde{H} N_j + 
\frac{1}{2} \mu_{ij} \; \overline{\psi^c_{iL}} \psi_{jL} + M_{ij} \overline{N_i} \psi_{jL} + {\rm h.c.}\, ,
\end{equation}
where $y^\nu$, $\mu$ and $M$ are $3\times 3$ complex matrices such
that their entries follow the hierarchy $\mu \ll y^\nu v \ll M$.  In
this case the active neutrino mass matrix takes the form $m_\nu \simeq
(v/\sqrt{2})^2 (y^\nu)^{T}(M^T)^{-1} \mu M^{-1} y^\nu$.  The mixing
between the light neutrino and one of the heavy states, on the other
hand, is not sensitive to $\mu$. In a simplified setting with only one
generation, the mixing would just be $y^\nu v/M$. Thus, the inverse
seesaw scenario clearly allows for sub-eV neutrinos masses, new
degrees of freedom not far from the weak scale and still a sizable
light-heavy mixing.  In order to have the right-handed fermions close
to the electroweak scale, the inverse seesaw scenario requires the
lepton number breaking parameter (matrix) $\mu$ to be at the keV
scale. Thus, the ``unnatural'' Yukawas are exchanged by the large
hierarchy $\mu/v\sim10^{-8}$. It can be argued that $\mu$ is naturally
small in the t'Hooft sense: setting $\mu\to 0$ restores a global
symmetry of the Lagrangian and therefore its renormalization group
running is only multiplicative.
 
 A completely different possibility to generate neutrino masses is
 given by neutrinophilic Two-Higgs-Doublet Models
 ($\nu$2HDM)~\cite{Gabriel:2006ns,Haba:2011nb,Davidson:2009ha}. In
 this framework, a symmetry, say $Z_2$ or $U(1)$, forces one of the
 Higgs doublets to couple to all SM fermions, thus being responsible
 for their masses, while the other Higgs couples to the lepton doublets
 and right-handed neutrinos.  This second Higgs would acquire a vacuum
 expectation value (vev) around the eV scale leading to naturally
 small neutrino masses. However, due to the smallness of the second
 vev, without any explicit breaking of the aforementioned symmetry,
 these models are either ruled out or considerably constrained by
 electroweak precision data~\cite{Machado:2015sha} and low energy
 flavor experiments~\cite{Bertuzzo:2015ada}.

The purpose of this work is to show how the idea of $\nu$2HDM can be
used to build a realistic, phenomenologically viable, and dynamical
inverse seesaw model.  This is achieved by promoting the $U(1)$
symmetry, used in Ref.~\cite{Davidson:2009ha} to avoid a Majorana mass
for the right-handed neutrinos, to a local symmetry.  Immediately,
chiral anomalies require the presence of additional fermion content
beyond the three right-handed neutrinos charged under this $U(1)$. A
minimal choice would be to double the spectrum of right-handed
neutrinos, with the additional ones having opposite $U(1)$
charges. Such minimal setup is identical to the inverse seesaw
framework.  In order to break this extra symmetry and give mass to
neutrinos, we introduce a new scalar particle, singlet under the SM
gauge group but charged under the additional $U(1)$, in such a way
that we dub our model $\nu$2HDSM (neutrinophilic Two-Higgs-Doublet +
Singlet Model), to distinguish it from the $\nu$2HDM already
considered in the literature. Other proposals for a UV completion of
the inverse seesaw mechanism can be found
in~\cite{1206.2590,1303.4887,1408.4785,1506.06946,1605.00239}.

The paper is organized as follows. In Section~\ref{sec:model} we build
the $\nu$2HDSM, analyzing in detail the scalar potential of the model, the
modified gauge Lagrangian and the neutrino
mass matrix. Section~\ref{sec:constraints} is devoted to the theoretical
and experimental constraints which we will consider, while we present
and discuss the results in Section~\ref{sec:results} and finally
conclude in Sec.~\ref{sec:conclusion}.

%%%%%%%%%%%%%%%%%%%%%%%%%%%%%%%%%%%%%%%%
\section{Neutrinophilic Two-Higgs-Doublet + Singlet Model}\label{sec:model} 
%%%%%%%%%%%%%%%%%%%%%%%%%%%%%%%%%%%%%%%%
%%
%%
\begin{table}
	\begin{center}
		\begin{tabular}{|c|ccc|}
		\hline
			& $SU(2)_L$ & $U(1)_Y$ & $U(1)_X$ \\
		\hline \hline
	    $\Phi_1$ & ${\bf 2}$ & $\frac{1}{2}$ & $\frac{1}{2}$ \\
	    $\Phi_2$ & ${\bf 2}$ & $\frac{1}{2}$ & 0 \\
	    $S$ & ${\bf 1}$ & 0 & $1$ \\
		\hline
	    $N_i^c$ &  ${\bf 1}$ & $0$ & $-\frac{1}{2}$\\
	    $\psi_{iL}$ & ${\bf 1}$ & $0$ & $\frac{1}{2}$\\\hline
		\end{tabular}
	\end{center}
	\caption{\label{tab:gauge_charges} Transformation properties of the scalar and fermion fields under $SU(2)_L \times U(1)_Y \times U(1)_X$.}
\end{table}
As explained in the Introduction, a global $U(1)_X$ symmetry (under
which the neutrinophilic Higgs doublet $\Phi_1$ and the right-handed
neutrinos $N_i$ have the same charge) is introduced in
Ref.~\cite{Davidson:2009ha}, allowing for a Yukawa interaction
$\overline{\ell}_{iL} \tilde{\Phi}_1 Y^\nu_{ij} N_j $ and forbidding a
$\overline{N}_i^c N_i$ Majorana mass term. In addition, it forbids the
other scalar doublet, $\Phi_2$, from coupling to neutrinos. We here
promote such $U(1)_X$ to a local symmetry. The first interesting fact
that we notice is that the gauging of $U(1)_X$ requires additional
fermion content to cancel anomalies.  Adopting minimality as a guide,
the obvious extension of the fermion sector is to add a second set of
three fermions $\psi_{iL}$, singlets under the SM local group but with
$U(1)_X$ charges opposite to $N_i^c$. This straightforward setup
already has most of the elements of the inverse seesaw scenario --
only the lepton number breaking parameter $\mu$ is missing.  The $\mu$
parameter can be generated dynamically by introducing a new scalar
degree of freedom, which we call $S$, charged only under $U(1)_X$. We
summarize all the charges (with the $U(1)_X$ ones conventionally
normalized to $\pm 1/2$) in Table~\ref{tab:gauge_charges}. The
relevant neutrino Lagrangian is
\begin{equation}\label{eq:neutrinos}
	-{\cal L}_\nu = \overline{\ell}_{iL} \tilde{\Phi}_1 Y^\nu_{ij} N_j + 
\frac{1}{2} S^* \overline{N^c_i} Y^N_{ij} N_j + \frac{1}{2} S^* \overline{\psi^c_{iL}} 
Y^\psi_{ij} \psi_{jL} + \overline{N_i} M_{ij} \psi_{jL} + {\rm h.c.}\, ,
\end{equation}
where $Y^\nu$, $Y^N$, $Y^\psi$ and $M$ are complex $3\times 3$ matrices.
We emphasize again that one of the consequences of gauging the
$U(1)_X$ symmetry is to introduce the typical particle content and the
Lagrangian of the inverse seesaw mechanism (although with an
additional Majorana mass term for the right-handed neutrinos $N_i$,
which plays a minor role in the neutrino mass mechanism). Before
studying the phenomenological consequences of the $\nu$2HDSM, let us
notice that there is an accidental $U(1)_\ell$ symmetry in the
Lagrangian, whose charges are given by
\begin{equation}
	U(1)_\ell \; \mbox{charges} \to \left\{ \Phi_1, \Phi_2, S, \ell_{iL}, N_i, \psi_{iL} \right\} = \left\{ 0,0,2q, q, q, q \right\} \, .
\end{equation}
This accidental symmetry corresponds to lepton number, extended in the
scalar sector only to $S$. Since the accidental symmetry
extends also to the scalar potential, we can already predict the
presence of a massless Nambu Goldstone Boson in the spectrum. This is
completely analog to what happens in Majoron
models~\cite{Chikashige:1980ui,Gelmini:1980re}, and we have explicitly
checked that the massless scalar is problematic for the model to pass
constraints from Electroweak Precision Measurements  and bounds
on Axion-Like Particles (ALPs)~\cite{1210.3196}. Pushing forward the
analogy with Majorons, we introduce an explicit breaking of the
accidental lepton number symmetry~\cite{Rothstein:1992rh} through a
dimension 5 term in the scalar potential
\begin{equation}\label{eq:symm_breaking}
	{\cal L}_{sb} = -\frac{g_{\rm{NP}}}{\Lambda}S(\Phi_1^{\dagger}\Phi_2)^2+{\rm{h.c.}}\, ,
\end{equation}
where $g_{NP}$ is a generic New Physics (NP) coupling and $\Lambda$
the scale associated with the breaking of the accidental symmetry. We
notice that, given the particle content and charges of our model, it is not
possible to write down $U(1)_\ell$ terms with dimension smaller than 5
involving only scalars. Moreover, the addition of explicit lepton
number breaking operators could, in principle, contribute to neutrino
masses. Nevertheless, the smallest $U(1)_\ell$ breaking term involving
fermions is given by the dimension 9 operator $(\overline{\ell_L}
\tilde{\Phi}_1) (\ell_L^c \tilde{\Phi}_1) (\Phi_2^\dag \Phi_1)^2$, and
thus these contributions are expected to be negligible.

It is well known that such breaking can be expected at least at the
Planck scale, but we keep open the possibility that $\Lambda \neq
M_{PL}$. We will come back to the value of $\Lambda$ in our
phenomenological analysis.

%%%%%%%%%%%%%%%%%%%%%%%%%%%%%%%%%%%%%%%%
%%%%%%%%%%%%%%%%%%%%%%%%%%%%%%%%%%%%%%%%
\subsection{Scalar Potential and Scalar Masses}\label{sec:potential} 
%%%%%%%%%%%%%%%%%%%%%%%%%%%%%%%%%%%%%%%%
%%%%%%%%%%%%%%%%%%%%%%%%%%%%%%%%%%%%%%%%
The most general $\nu$2HDSM potential compatible with the
$SU(2)_L\times U(1)_Y\times U(1)_X$ charges of
Table~\ref{tab:gauge_charges} with the addition of the only dimension 5 operator that breaks $U(1)_\ell$ is the following:
\begin{eqnarray}\label{potential}
\mathcal{V}\big(\Phi_1,\Phi_2,S\big)&=&m_{11}^2\Phi_1^{\dagger}\Phi_1+m_{22}^2\Phi_2^{\dagger}\Phi_2\nonumber\\
&+&\frac{\lambda_1}{2}\big(\Phi_1^{\dagger}\Phi_1\big)^2+\frac{\lambda_2}{2}\big(\Phi_2^{\dagger}\Phi_2\big)^2+\lambda_3\big(\Phi_1^{\dagger}\Phi_1\big)\big(\Phi_2^{\dagger}\Phi_2\big)+\lambda_4\big(\Phi_1^{\dagger}\Phi_2\big)\big(\Phi_2^{\dagger}\Phi_1\big)\nonumber\\
&+&m_{S}^2|S|^2+\frac{\lambda_s}{2}|S|^4+\big(\lambda_{1s}\Phi_1^{\dagger}\Phi_1+\lambda_{2s}\Phi_2^{\dagger}\Phi_2\big)|S|^2 +\left[\frac{g_{\rm{NP}}}{\Lambda}S(\Phi_1^{\dagger}\Phi_2)+{\rm{h.c.}}\right]\, ,
\end{eqnarray}
whose minimization is shown in Appendix~\ref{app:details}. To fix the
notation, we explicitly write the scalar fields as
\begin{equation}\label{doublets}
\Phi_{1,2}=
\begin{pmatrix}
\phi_{1,2}^+\\
\frac{1}{\sqrt{2}}\big(v_{1,2}+\rho_{1,2}+i\eta_{1,2}\big)\\
\end{pmatrix},~~~~~~ S=\frac{1}{\sqrt{2}}(v_3+\rho_3+i\eta_3)\, ,
\end{equation}
where $v_{1,2,3}$ are the corresponding vevs of $\Phi_{1,2}$ and $S$,
the EW scale $v=\sqrt{v_1^2+v_2^2}$ and we define $t_\beta=\tan\beta
\equiv v_2/v_1$, so that $c_\beta=\cos \beta \equiv v_1/v$ and $s_\beta=\sin
\beta \equiv v_2/v$. After spontaneous symmetry breaking, two charged and two
neutral would-be Nambu-Goldstone bosons are absorbed by the $W^{\pm}$,
$Z$ and a new gauge boson $X$. The remaining particle spectrum
consists of two charged ($H^{\pm}$) and four neutral scalar fields:
three CP-even $(h,H,s)$ and one CP-odd ($A$). The CP-odd scalar would
be the massless Majoron in the absence of the explicit $U(1)_\ell$
breaking term ${\cal L}_{sb}$ in Eq.~(\ref{eq:symm_breaking}). The
detailed analysis of the scalar eigenstates is presented in
Appendix~\ref{app:details}.

Note that Higgs measurements at the LHC require one of the CP-even
eigenstates, $h$, to be sufficiently close to the SM-like Higgs
$h_{SM}= c_\beta \, \rho_1 + s_\beta \, \rho_2$. We can write the scalar
mass eigenstates approximately as
\begin{equation}\label{rotations}
\begin{array}{rcl}
H^{\pm} & = & \ds{-s_\beta  \; \phi_1^{\pm}+c_\beta  \; \phi_2^{\pm}\, ,}\\
h & \simeq & \ds{c_\alpha \, h_{SM} - s_{\alpha} \, H_{SM} + U_{31} \, \rho_3 \, ,}\\
H & \simeq & \ds{s_\alpha \, h_{SM} + c_{\alpha} \, H_{SM} + U_{32} \, \rho_3 \, ,}\\
s & \simeq & \ds{U_{13} \, h_{SM} + U_{23} \, H_{SM} + \rho_3 \, , }\\
A & = &\frac{v_1 v_2}{\sqrt{v_1^2 v_2^2+4 v^2 v_3^2}}(\ds{\frac{2v_3}{v_1}\, \eta_1-\frac{2v_3}{v_2}\, \eta_2+\eta_3)\, .}
	\end{array}
\end{equation}
Here, $H_{SM}$ is the combination orthogonal to $h_{SM}$, $c_\alpha=\cos\alpha$, $s_\alpha=\sin \alpha$
parametrize most of the mixing and $U_{ij}$ encode the rest of the mixing matrix. 
The alignment limit is defined by $h=h_{SM}$, which requires $s_\alpha=U_{13}=U_{31}=0$.
The corresponding masses are given by
\begin{equation}\label{scalarMasses}
	\begin{array}{lcl}
m_{H^{\pm}}^2&=&\ds{-\frac{\lambda_4\Lambda+\sqrt{2} \, g_{\rm{NP}} v_3 }{2\Lambda}v^2 \, ,}\\
m_{h, H}^2 & \simeq & \ds{\frac{1}{2} \left( \lambda_1 v_1^2 + \lambda_2 v_2^2  \pm \sqrt{(\lambda_1 v_1^2 - \lambda_2 v_2^2)^2 + 4 \lambda_{34}^2 v_1^2 v_2^2+\frac{8v_1^2 v_2^2}{\Lambda^2}(v_3^2-\sqrt{2}\lambda^2_{34}v_3\Lambda)} \right) \, ,}\\
m_s^2&\simeq& \ds{\lambda_s v_3^2 -\frac{g_{\rm{NP}}}{2\sqrt{2}\Lambda}\frac{v_1^2v_2^2}{v_3} \, ,}\\
m_A^2&=& \ds{\frac{-g_{\rm{NP}}}{2\sqrt{2}\Lambda}\frac{(v_1^2v_2^2+4v^2v_3^2)}{v_3} \, ,}
\end{array}
\end{equation}
where $\lambda_{34}=\lambda_3+\lambda_4$.  Notice that we need
$\lambda_4\Lambda+\sqrt{2} \, g_{\rm{NP}} v_3<0$ for electromagnetism
not to be broken, and $g_{NP}<0$ to ensure a positive mass for $A$. In
what follows, we will always assume this to be the case. In fact, as
we will see later requiring $\lambda_4\Lambda+\sqrt{2} \, g_{\rm{NP}}
v_3<0$ is not a problem for the stability of the potential.

%%%%%%%%%%%%%%%%%%%%%%%%%%%
%%%%%%%%%%%%%%%%%%%%%%%%%%%
\subsection{Gauge Lagrangian}\label{sec:gauge}
%%%%%%%%%%%%%%%%%%%%%%%%%%%
%%%%%%%%%%%%%%%%%%%%%%%%%%%
Let us now move to the analysis of the gauge bosons and their masses.
In what follows, we will, for simplicity, assume that there is no
$B_{\mu\nu}X^0_{\mu\nu}$ kinetic mixing (we call $X^0_\mu$ the gauge
boson associated with $U(1)_X$). The  scalar kinetic terms are 
\begin{eqnarray}\label{gaugeLagrangian}
\mathcal{L}_G=\big(D_\mu\Phi_1\big)^\dagger\big({D}^\mu\Phi_1\big)+\big(D_\mu\Phi_2\big)^\dagger\big({D}^\mu\Phi_2\big)+\big(D_\mu S\big)^\dagger\big({D}^\mu S\big)\, ,
\end{eqnarray}
with the covariant derivatives given in Eq.~(\ref{covariant}). The
masses of the charged gauge bosons are easily computed and are
identical to their values in the SM. In the neutral sector, instead,
we have mixing between the $\{ W^3_\mu, B_\mu , X^0_\mu \}$ states,
resulting in the mass matrix given in Eq.~(\ref{neutralGauge}). Apart
from the massless photon, the $Z$ and $X$ boson masses are given by
\begin{equation}\label{eq:gauge_masses}
m^2_{Z,X}=\frac{v^2}{8}\Big[g^2+{g^\prime}^2+g_X^2 \,b^2% c_{\beta{^\prime}}^2
\pm\sqrt{(g^2+{g^\prime}^2-g_X^2 \, b^2%c_{\beta^{\prime}}^2
)^2+
4(g^2+{g^\prime}^2)g_X^2 \, c_\beta^4}\Big]\, ,
\end{equation}
where $g$, $g^\prime$ and $g_X$ are, respectively, the $SU(2)_L$,
$U(1)_Y$ and $U(1)_X$ coupling constants and $b\equiv \sqrt{v_1^2+4v_3^2}/v$. As we will see later, the electroweak 
$\rho=\frac{m^2_W}{c_W^2m_Z^2}$ parameter will generically require
$g_X v_1\ll g v$. 
 In this limit, we find that the $Z-X$ mass squared matrix becomes
\begin{eqnarray}\label{XZmassmatrix}
\mathcal{M}^2_{ZX}=
\frac{1}{4}
\begin{pmatrix}
(g^2+{g^{\prime}}^2) v^2& -\sqrt{g^2+{g^{\prime}}^2} g_X v_1^2\\
-\sqrt{g^2+{g^{\prime}}^2} g_X v_1^2 & g_X^2(v_1^2+4v_3^2)
\end{pmatrix}\simeq
\begin{pmatrix}
m_{Z}^2& - m_{Z} m_{X} \delta \\
-m_{Z} m_{X} \delta & m_{X}^2
\end{pmatrix}
\, ,
\end{eqnarray}
with the approximate masses
\begin{equation}\label{gaugeMasses}
	\begin{array}{rcl}
		m_Z^2 &  \simeq & \ds{\displaystyle \frac{1}{4}(g^2+{g^{\prime}}^2)\, v^2\, ,} \\
		m_X^2 & \simeq & \ds{\displaystyle \frac{1}{4}g_X^2(v_1^2+4v_3^2) \,  ,}
	\end{array}
\end{equation}
and the $Z-X$ mass mixing parameter $\delta$ given by
\begin{eqnarray}\label{delta}
\delta=\frac{v_1^2}{v\sqrt{v_1^2+4v_3^2}}.
\end{eqnarray}
The mass eigenstates in this case follow
\begin{equation}\label{neutralbosons}
	\begin{array}{rcl}
		W^3_\mu & \simeq & c_W \, Z_\mu+s_W \, A_\mu+g_X \, \dfrac{c_\beta^2}{g}c^2_W \, X_\mu \, ,\\
		B_\mu & \simeq & -s_W\, Z_\mu+c_W \, A_\mu-g_X\, \dfrac{c_\beta^2}{g}c_W s_W \,X_\mu \, ,\\
		X^0_\mu & \simeq &-g_X\dfrac{c_\beta^2}{g}c_W \, Z_\mu+X_\mu \, ,
	\end{array}
\end{equation}
where $c_W=\cos \theta_W$, $s_W=\sin \theta_W$ and
$g_X\frac{c_\beta^2}{g}c_W = \frac{g_X
  v_1^2}{v^2\sqrt{g^2+{g^{\prime}}^2}}\ll 1$ is approximately the
$X-Z$ mixing, its exact expression is given in Eq.~(\ref{ct}).  Due to
this mixing with the $Z$, the new gauge boson will acquire chiral
couplings to all SM fermions. The coupling of $X$ to a
fermion $f$ is given by
\begin{equation}\label{gXff}
  g_{X\bar{f}{f}}=g_X\frac{v_1^2}{v^2}(I_3^f-s_W^2Q_f),
\end{equation} 
where $I_3^f=0,\pm1/2$ and $Q_f$ are, respectively, the isospin and electric charge of $f$.
%%%%%%%%%%%%%%%%%%%%%%%%%%%
%%%%%%%%%%%%%%%%%%%%%%%%%%%
\subsection{The Neutrino Sector}\label{sec:neutrino}
%%%%%%%%%%%%%%%%%%%%%%%%%%%
%%%%%%%%%%%%%%%%%%%%%%%%%%%
Let us now go back to the consequences of Eq.~(\ref{eq:neutrinos}). As
already stressed, what we have obtained is a generalization of the
inverse seesaw Lagrangian, with an additional Majorana mass term for
the right-handed neutrinos $N_i$. In the $(\nu_{iL}, N_i^c,
\psi_{iL})$ basis, the $9\times 9$ mass matrix is
\begin{equation}\label{bigM}
{\cal M} = \left( 
\begin{array}{ccc}
0 & Y^\nu v_1  & 0 \\
Y^{\nu T} v_1  &  Y^N v_3 & M \\
0 & M^T &  Y^\psi v_3
\end{array}
\right) \, .
\end{equation}
Comparing with Eq.~(\ref{eq:inv_seesaw}), we see that we obtain $\mu =
Y^\psi v_3 $, with the hierarchy among the elements of the matrices
$Y^N v_3 , Y^\psi v_3 \ll Y^\nu v_1$ automatically satisfied if $Y^N ,
Y^\psi \ll Y^\nu$ (assuming $v_1$ and $v_3$ of the same order). If we
further assume the hierarchy $Y^\nu v_1 \ll M$, the lower right block
of ${\cal M}$ will have eigenvalues of the order of the eigenvalues of
$M$, allowing us to integrate out $N_i$ and $\psi_{iL}$. The first
terms in a $\inv{M}$ expansion of the neutrino mass matrix are then
\begin{align}\label{eq:neutr_mass}
  m_\nu \simeq\, &(Y^\nu v_1) \inv{M^T} (Y^\psi v_3 ) \inv{M} (Y^{\nu T} v_1) \\
  &+ (Y^\nu v_1) \inv{M^T} (Y^\psi v_3) \inv{M} (Y^N v_3 ) \inv{M^T} ( Y^\psi v_3) \inv{M} (Y^{\nu T} v_1) \, , \nonumber
\end{align}
from which we obtain the inverse seesaw contribution (the first term)
plus a small correction dependent on $Y^N$. Notice that the neutrino
masses vanish in the $v_3 \to 0$ or $Y^\psi \to 0$ limit, but not for
$Y^N \to 0$. There are many parameters in the neutrino sector which
are independent from the parameters in the scalar and gauge sector,
making it simple to reproduce the observed values of the neutrino
oscillation parameters in relevant portions of the parameter space, as
we checked explicitly. Besides, this scenario could comprise
leptogenesis and a WIMP-like dark matter candidate, but we do not
pursue such possibilities in this manuscript.

For simplicity, we take $M$ to be at the TeV scale,
  leading to TeV heavy neutrinos. The phenomenology of our inverse
  seesaw scenario may differ from the usual one due to the presence of the
  new gauge boson and the extended scalar sector, but we do not study
  it in this manuscript. As a final comment regarding this subject
  we note that if the mass scale $M$ is assumed to be lower, e.g., a few GeV, $Z$ 
  and $h$ may, through mixing, decay to heavy neutrinos. 
  The $Z-X$ mixing could be suppressed by having a large
  $v_3$ (see Eq.~(\ref{delta})) and the $h-S$ mixing could also be suppressed by small Yukawas, 
  so both decays can be made negligible.

%%%%%%%%%%%%%%%%%%%%%%%%%%%
%%%%%%%%%%%%%%%%%%%%%%%%%%%
\section{Theoretical and Experimental Constraints}\label{sec:constraints}
%%%%%%%%%%%%%%%%%%%%%%%%%%%
%%%%%%%%%%%%%%%%%%%%%%%%%%%
Let us now list and explain the constraints we will impose on the
parameter space of the $\nu$2HDSM. In the numerical analysis of
Section~\ref{sec:results} these constraints will be imposed to assess
the phenomenological viability of the model. For clarity, we will
distinguish between theoretical and experimental constraints.

\noindent{\bf \uline{Theoretical Constraints}: }
\begin{description}
\item[Perturbativity] As a simplified approach, to ensure tree level perturbativity, we only require the absolute value of all the quartic couplings to be smaller than $4\pi$, namely
\begin{equation}\label{eq:perturbativity}
|\lambda_{1}|,~|\lambda_{2}|,~|\lambda_{3}|,~|\lambda_{4}|,~|\lambda_{s}|,~|\lambda_{1s}|,~|\lambda_{2s}|<4\pi.
\end{equation}

\item[Vacuum Stability] To have a potential bounded from below, the quartic couplings need to satisfy some stability conditions at tree level~\cite{Drozd:2014yla}, 
\begin{equation}\label{eq:vac_stab}
	\begin{array}{c}
		\lambda_{1,2,s}>0 \, ,~~~~~\lambda_3>-\sqrt{\lambda_1\lambda_2} \, ,~~~~~\lambda_{34}>-\sqrt{\lambda_1\lambda_2} \, ,\\
		\lambda_{1s}>-\sqrt{\lambda_1\lambda_s} \, ,~~~~~\lambda_{2s}>-\sqrt{\lambda_2\lambda_s} \, .
	\end{array}
\end{equation}
As we already stressed, having $\lambda_4\Lambda+\sqrt{2} \, g_{\rm{NP}} v_3 <0$, in order to guarantee an unbroken $U(1)_{\rm em}$, is not in contradiction with a stable potential.

\item[Local minima for the potential] To be certain that the vevs in
  our model are local minima of the potential, we impose all the tree
  level masses of Eq.~(\ref{scalarMasses}) to be positive. 
\end{description}

%%%%%%%%%%%%%%%%%%%%%%%%%%%
\noindent{\bf \uline{Experimental Constraints}:}
%%%%%%%%%%%%%%%%%%%%%%%%%%%
\begin{description}
%%%%%%%%%%
\item[Electroweak Precision Measurements (EWPM)]
%%%%%%%%%%
As is well known, EWPM are crucial in assessing the validity of a
model.  However, in our analysis we will make some simplifying
assumptions which we believe will capture the essential bounds. First
of all, the usual treatment in terms of oblique
parameters~\cite{Peskin:1991sw} is valid only when a mass gap is
present between the EW and the NP scales. As it is clear from the
masses we are considering,
Eqs.~(\ref{scalarMasses})-(\ref{eq:gauge_masses}), this is not the
case in the $\nu$2HDSM. In principle, a better analysis would involve
an extended set of oblique
parameters~\cite{Maksymyk:1993zm,Burgess:1993mg}. For simplicity, we
will just perform the analysis in term of the usual Peskin-Takeuchi
self-energy parameters $S$, $T$ and $U$.  In addition, since we always
assume the $U(1)_X$ gauge coupling $g_X$ to be small to ensure $\rho
\simeq 1$ already at tree level (see Eq.~(\ref{eq:gauge_masses})), and
since all the $X_\mu$ contributions to the gauge bosons vacuum
polarization corrections to the EW precision observables are
suppressed at least by a factor $g_X^2$, we will neglect such
contributions. We have explicitly checked that the loops we are
neglecting are small as expected. In practice, we compute the oblique
parameters using a Two-Higgs-Doublet+Singlet
approximation~\cite{Grimus:2008nb}, imposing compatibility at 1, 2 and
3 $\sigma$ with the following fitted values~\cite{Baak:2014ora}
\begin{eqnarray}\label{eq:STU}
\Big(S^{\rm{fit}},T^{\rm{fit}},U^{\rm{fit}}\Big)=\Big(0.05\pm0.11,0.09\pm0.13,0.01\pm0.11\Big).
\end{eqnarray}
%
%%%%%%%%%%

%%%%%%%%%%
\item[Higgs decays]
%%%%%%%%%%
Since several neutral particles with masses smaller than half of the SM
Higgs mass may exist in our model, we can have new contributions to
the Higgs invisible width. To be conservative, we will take the rather
aggressive bound ${\rm{BR}}(h\to{\rm{invisible}}) \lesssim 0.13$ at
$95\%$ C.L.~\cite{Ellis:2013lra}, and $\Gamma_{\rm tot}^{\rm SM} =
4.07$ MeV as the SM Higgs total decay width~\cite{Denner:2011mq}. The
important invisible channels in our model are the following:
\begin{description}
	\item[\uline{$h\to \mathcal{S}\mathcal{S}$}] (where
          $\mathcal{S}=H,s,A$),
          with decay rate
		\begin{equation}\label{eq:htoSS}
				\Gamma(h\to\mathcal{S}\mathcal{S}) = \frac{|g_{h\mathcal{S}\mathcal{S}}|^2}{32\pi m_h}\sqrt{1-\frac{4m^2_{\mathcal{S}}}{m^2_h}}\, .
		\end{equation}
		
		The expressions for the $g_{h\mathcal{S}\mathcal{S}}$ couplings above are very lengthy and not much illuminating, and so we do not write them here.
	\item[\uline{$h\to XX$}] with decay width 
		\begin{eqnarray}\label{hXX}
			\Gamma(h\to XX)=\frac{|g_{hXX}|^2 }{64\pi}\frac{m_h^3}{ m_X^4 } \sqrt{1-\frac{4m_X^2}{m_h^2}} \left[ 1 - \frac{4m_X^2}{m_h^2}+12\frac{m_X^4}{m_h^4}  \right] \, .
		\end{eqnarray}
		Assuming $g_X \ll 1$ and using Eqs. (\ref{eq:CPeven_true_rotation}) and (\ref{eq:CPeven_true_rotation2}) the coupling becomes
				\begin{eqnarray}\label{g_hXX}
		g_{hXX}=\frac{g_X^2}{4v}[v_1^2 c_\alpha +v_1 v_2 s_\alpha +4 v v_3 U_{31}]\,.
				\end{eqnarray}

	\item[\uline{$h\to ZX$}] with decay width 
		\begin{eqnarray}\label{hZX}
		\Gamma(h\to ZX)&=&\frac{|g_{hZX}|^2}{64\pi  }\frac{ \left[ (m_h^2-m_Z^2)^2 +m_X^2(m_X^2+10m_Z^2-2m_h^2)\right]}{m_h m_X^2 m_Z^2}\nonumber\\
		&\times& \sqrt{1-\frac{(m_X+m_Z)^2}{m_h^2}}\sqrt{1-\frac{(m_X-m_Z)^2}{m_h^2}} \, 
		\end{eqnarray}
		where the coupling, in the limit $g_X\ll 1$, reads
			\begin{eqnarray}\label{g_hZX}
				g_{hZX}=-\frac{g g_X}{2c_W}v_1 s_\alpha\,.
			\end{eqnarray}
	\item[\uline{$h\to A\mathcal{V}$}] (where $\mathcal{V}=X,Z$), with decay rate
		\begin{eqnarray}\label{hAV}
			\Gamma(h\to A\mathcal{V})= \frac{|g_{hA\mathcal{V}}|^2 }{16 \pi} \frac{m_h^3}{m_\mathcal{V}^2} \left[ 1-\frac{(m_\mathcal{V}-m_A)^2}{m_h^2} \right]^{3/2}  \left[ 1-\frac{(m_\mathcal{V}+m_A)^2}{m_h^2} \right]^{3/2} \, ,
		\end{eqnarray}
		and couplings 
		\begin{eqnarray}\label{g:hAV}		
			g_{hAX}&=&\frac{g_X[-U_{31} v_1+\cos(\alpha-\beta)v_3]}{\sqrt{v_1^2+4v_3^2}}\\
			g_{hAZ}&=&-\frac{\sqrt{g^2+{g^{\prime}}^2}}{\sqrt{v_1^2+4v_3^2}} v_3 s_\alpha\,.
		\end{eqnarray}

\end{description}

In addition to the constraint on the Higgs invisible decay width, we
must verify that the visible decay widths of $h$ into SM particles are
not too different from their SM values.  We thus have
\begin{description}
\item[\uline{$h \to SM \, SM$}] where $SM$ stands for all the SM channels measured at the LHC. According to~\cite{1606.02266}, we allow for at most a $15\%$ departure from the SM values. 
\end{description}

\item[Z invisible width] The decay of $Z$ to light particles through
  $Z\to \mathcal{S} A$ and $Z \to X {\cal S}$ (where
  $\mathcal{S}=H,s)$ can contribute to the $Z$ invisible width. Comparing
  the LEP result with  the SM prediction for $Z$ to invisible, we find
  that $\Gamma^{\rm{NP}}(Z\to{\rm{invisible}})<1.8~$MeV at $3\sigma$~\cite{Machado:2015sha}. The partial decay width for such processes are
\begin{eqnarray}\label{GammaZ}
	\begin{array}{rcl}
		\Gamma(Z\to \mathcal{S}A) & =&  \ds{\frac{|g_{Z\mathcal{S}A}|^2}{16\pi} m_Z \left[ 1-\frac{(m_{\cal S}-m_A)^2}{m_Z^2} \right]^{3/2}  \left[ 1-\frac{(m_{\cal S}+m_A)^2}{m_Z^2} \right]^{3/2} \, ,}\\ \vspace{0.3cm}
		\Gamma(Z \to {\cal S} X) & =& \ds{\frac{|g_{Z  {\cal S} X}|^2}{64\pi  }\frac{ \left[ (m_Z^2-m_ {\cal S}^2)^2 +m_X^2(m_X^2+10m_Z^2-2m_{\cal S}^2)\right]}{m_X^2 m_Z^3} } \\
		&\times& \ds{\sqrt{1-\frac{(m_X+m_{\cal S})^2}{m_Z^2}}\sqrt{1-\frac{(m_X-m_{\cal S})^2}{m_{\cal S}^2}} \, ,}
	\end{array}
\end{eqnarray}
	where, assuming $g_X\ll 1$ and using
        Eqs. (\ref{eq:CPeven_true_rotation}) and
        (\ref{eq:CPeven_true_rotation2}), the couplings become
\begin{eqnarray}\label{Zcouplings}
g_{ZHA}&=& -\frac{g v_3 c_\alpha}{c_W\sqrt{v_1^2+4v_3^2}}\,,\\
g_{ZsA}&=& -\frac{gv_3 (U_{23}- c_\beta U_{13})}{c_W\sqrt{v_1^2+4v_3^2}}\,,\\
g_{ZHX}&=& -\frac{gg_X v_1 c_\alpha}{2c_W}\,,\\
g_{ZsX}&=& -\frac{gg_X v_1 (U_{23}-c_\beta U_{13})}{2c_W}\,.
\end{eqnarray}
\item[Lower bound on the charged Higgs mass]
The LEP II experiment has searched for the double production of
charged Higgs bosons in events with center-of-mass energy from $183$
GeV to $209$ GeV, with a total luminosity of $2.6$ fb$^{-1}$
\cite{Abbiendi:2013hk}. Since no excess was found, we have a lower
bound on the charged Higgs boson mass
\begin{equation}
	m_{H^\pm} \gtrsim 80\,{\rm GeV} .
\end{equation}
In principle, bounds from the LHC should also be considered. These
bounds depend, however, on $t_\beta$.  In contrast, the LEP bound is
obtained using charged Higgses pair production via a $Z$ boson in the
s-channel, which is $t_\beta$ independent.  In our model, in the large
$t_\beta$ limit, we have $H^\pm \to \phi_1^\pm$, which does not couple
to quarks, so in this case the LEP limit prevails.  In the region
$t_\beta \sim 1$ the LHC bounds can be stronger
than the LEP limit (in particular, if the channel $H^\pm \to \ell^\pm + N$ is open, the decay is similar to the one of sleptons in the R-parity conserving MSSM, and the bound would be $m_{H^\pm} \gtrsim 300$ GeV~\cite{Khachatryan:2014qwa}). Nevertheless, a dedicated analysis would be needed
to compute the correct limit in this case and this is outside the
scope of our paper. Moreover, as we will see in what follows, this
region will be excluded by other bounds.

\item[Bounds on the pseudoscalar mass and couplings]
Depending on the size of the lepton number breaking term of Eq. (\ref{eq:symm_breaking}), a light pseudoscalar may be present in the spectrum. Such $A$ couples to electrons through its $\eta_2$ component (see Eq. (\ref{rotations})), we get a coupling
\begin{equation}
	g_{Aee} = \frac{2 v_3 m_e}{v  t_\beta \sqrt{v^2 s^2_{2\beta} + 4 v_3^2}} .
\end{equation}
For masses $m_A \lesssim 100$ keV, stellar cooling gives a strong bound
$g_{Aee} \lesssim 2.5 \times 10^{-13}$ \cite{1210.3196}. In our case,
this translates on the bound $v_3 \lesssim 10^{-5}$ GeV. This region of
parameter space is problematic because $\Gamma(h \to XX)$ becomes too
large. This can be seen using Eqs. (\ref{gaugeMasses}) and
(\ref{g_hXX}) in Eq. (\ref{hXX}), from which in the alignment limit
one gets $\Gamma(h\to XX) \simeq m_h^3/(64 \pi v^2)$, far above the
experimental limit. In order to avoid all the ALPs bounds which limit $v_3$, we will thus 
impose $m_A \gtrsim 10$ GeV  \cite{1210.3196}.
\item[Kaon  and $B$ decays]
%%%%%%%%%%
The presence of the light gauge boson $X$ that mixes with $Z$ can
greatly enhance the $K\to\pi+{\rm
  invisible}$~\cite{Anisimovsky:2004hr, Artamonov:2008qb} branching
ratio via loops~\cite{Davoudiasl:2014kua}.  There are various
experiments which constrain the $Z-X$ mass mixing parameter $\delta$
given in Eq.~(\ref{delta}) with the typical bound of
$|\delta|\lesssim10^{-3}-10^{-2}$ \cite{Davoudiasl:2012ag}. When
$v_3\ll v_1\ll v_2$, this parameter reads $\delta\simeq
v_1/v=c_\beta$. Recasting the analysis of
Ref.~\cite{Davoudiasl:2014kua} onto our scenario, we find that
$K\to\pi+{\rm invisible}$ constrains $c_\beta\lesssim 10^{-3}$, with
some mild dependence on the mass of the charged scalars. Therefore,
unless the decay is not kinematically accessible (or 114
MeV$<M_X<151$~MeV due to experimental cuts), kaon physics would force
$v_1\lesssim 100$~MeV, essentially ruling out the model.  Increasing
$v_3$ suppresses the mass mixing parameter, so that in the decoupling
limit, where $v_3\rightarrow \infty $, it goes to zero and the
constraint is satisfied automatically. \\ 
The $B\to K X$ process is akin to $K \to \pi X$. By performing a similar loop calculation
substituting $X$ by its associated Goldstone boson (making use of the
equivalence theorem), in the limit where all quarks but the top are
massless, we reach similar conclusions: unless the decay is not
kinematically accessible, the $B \to K \nu\nu$ branching ratio would
put a bound on $v_3 \gtrsim 10$~TeV or $t_\beta\ll 1$. Moreover,
there is some dependence on the mass of the charged Higgs boson. In
Fig.~\ref{fig:moneyplot}, we show the constraints for both kaon and
$B$ decays, for different values of $v_3$ and fixing
$M_{H^\pm}=500$~GeV.  We will see shortly that our model is viable in
the range $v_3\sim (10 \div 10^3)$ GeV.
%%%%%%%%%%
\item[$b\to s$ exclusive decays]
%%%%%%%%%%
If NP beyond the SM exists in the flavor sector, flavor changing
neutral current (FCNC) processes could be modified by the exchange of
unknown virtual particles. The rates of the $\bar{B}\to X_s
\gamma$ \cite{Misiak:2017bgg} and $B\to \ell^+\ell^-$
\cite{Arnan:2017lxi} channels, via the exchange of $H^\pm$ in a
radiative penguin diagram, put constraints on $m_{H^{\pm}}$ and
$t_\beta$ in our model. In particular, these processes set the lower
bound $t_\beta \gtrsim 0.8$ for $m_{H^\pm} \lesssim 600$ GeV.

%%%%%%%%%%
\item[Bounds on direct $X$ production from accelerator experiments]
%%%%%%%%%%
The coupling of $X$ to the SM charged fermions, via $Z-X$ mixing, can
be constrained by electron and proton beam-dump experiments and $B$-factories,  
similarly to what happens in the dark photon scenario
\cite{Bossi:2013lxa}. Such bounds are expected to depend strongly on
$t_\beta$ and $v_3$, since these parameters control the gauge boson
mixing (see Eqs.~(14) and (16)). At fixed target experiments $X$ could be 
produced by radiation or meson decays ($\pi^0\to\gamma X$) and 
subsequently decay to a $e^+ e^-$ pair that could be measured at 
the detector. In our model, the presence of the invisible $X$ partial width 
naturally reduces the $e^+ e^-$ branching ratio to $\sim14\%$, weakening this  
bound with respect to the dark photon case. At $B$-factories direct $X$ production could 
also be achieved, via the $X-Z$ mixing, through the process $e^+ e^- \to X
\mu^+\mu^-$. Moreover, differently from the dark photon phenomenology,
the $X$ axial coupling to fermions can induce $\Upsilon\to\gamma
X\to\gamma+{\rm invisibles}$ decays~\cite{Babu:2017olk}. Experimental
data bounds this branching ratio to be below $4.5\times 10^{-6}$.

\end{description}

%%%%%%%%%%%%%%%%%%%%%%%%%%%
%%%%%%%%%%%%%%%%%%%%%%%%%%%
\section{Analysis of our model}
\label{sec:results}
Let us now analyse the viability of the $\nu$2HDSM. We start by listing
the bounds that the experimental limits discussed in Section
\ref{sec:constraints} impose on the parameter space. First of all,
requiring $m_A \gtrsim 10$ GeV and $m_{H^\pm} \gtrsim 80$ GeV in
Eq.~(\ref{scalarMasses}) requires $g_{NP} < 0$ and $\lambda_4 -
\frac{\sqrt{2} v_3}{\Lambda} \lesssim -0.21$. The last condition
automatically ensures a positive charged Higgs mass, {\it i.e.}
unbroken electromagnetism. As for the lepton number breaking coupling,
for definitiveness we will fix it to $g_{NP} = -1$ from now on. The
requirement of being close to the alignment limit can be computed
rotating Eq. (\ref{Mrho}) to the Higgs basis. The conditions read
\begin{equation}\label{eq:alignment}
\lambda_{34} \simeq \frac{\lambda_1 + \lambda_2}{2} + \frac{\lambda_1 - \lambda_2}{2 c_{2\beta}} - \frac{\sqrt{2} g_{NP} v_3}{\Lambda}, ~~~\lambda_{1s} \simeq \lambda_{2s} \simeq - \frac{g_{NP} v^2 s_{2\beta}}{2 \sqrt{2} v_3 \Lambda} .
\end{equation}
Moving to the Higgs and $Z$ invisible decays, we first notice that
$\Gamma(Z \to HA)$ is always larger than the experimental limit when
this channel is open. We thus require to always have $m_A + m_H > m_Z$
to kinematically close the decay. The Higgs decays to scalars of
Eq.~(\ref{eq:htoSS}) are below the experimental limit when
$|g_{h{\cal SS}}| \lesssim 2.75$ GeV. While this is always true for
${\cal S} = s$, when ${\cal S}=H$ or $A$ this gives the bound
\begin{equation}\label{eq:constr_lambda34}
	\left| \lambda_{34} \pm \frac{\sqrt{2}v_3}{\Lambda}  \right| \lesssim 0.1 ,
\end{equation}
which applies when the channels are open. On the other hand, it is
simple to check that close to the alignment limit we have $\Gamma(h
\to XX) \simeq \frac{m_h^3}{64\pi v^2} \frac{c_\alpha^2 v_1^4}{(v_1^2
  + 4 v_3^2)^2}$, which can be suppressed by $v_1 \ll v_3$. The same
condition can also suppress the $h \to ZX$ and $Z \to HX$ decays, as
can be seen from the $m_X$ dependence in the denominator of Eqs.~(\ref{g_hZX}) and (\ref{Zcouplings}), while the $h\to AX$ and $h \to
AZ$ decays are always suppressed and unimportant.  It should be noted
that when the $h \to XX$, $h\to ZX$ and $Z\to HX$ channels are
kinematically closed, which happens when $m_X>m_h/2$ and $m_H\gtrsim
30~$GeV, the $v_1 \ll v_3$ constraint is removed and $v_1$ can assume 
larger values (as far as $t_\beta \geq 0.8$).

An immediate consequence of the requirement $v_1 \ll v_3$ is that the
pseudoscalar mass in this limit is given by $m_A^2 \simeq \frac{2
  v^2}{\sqrt{2}}\frac{v_3}{\Lambda}$, in such a way that the
requirement $m_A \gtrsim 10$ GeV implies $v_3/\Lambda \gtrsim
10^{-3}$. For definitiveness, in what follows we will fix $\Lambda =
10$ TeV, although other values may be allowed.

With these constraints in mind, we can now scan over the parameter
space of the model to assess if it is phenomenologically viable. To
this end, we randomly generate $18$ million points and impose all the
bounds described in Sec.~\ref{sec:constraints}. In addition, we also
require the Higgs boson mass to be in the range $123\, {\rm GeV} \leq
m_h \leq 127\, {\rm GeV}$, the Z boson mass to be within 3 $\sigma$ of
the LEP measured value, $m_Z = 91.1876 \pm 0.0021$
GeV~\cite{Olive:2016xmw} and we require to be close to the alignment
limit of Eq.~(\ref{eq:alignment}). More precisely, we perform a
linear scan over the range of the seven quartic couplings in
Eq.~(\ref{eq:perturbativity}), and a logarithmic scan over
$t_\beta$, $v_{3}$ and $g_X$ in the window
\begin{equation}\label{eq:range_parameters}
	0.8 \leq t_\beta \leq 246 \, , ~~~ 10\, {\rm GeV} \leq v_3 \leq 1\, {\rm TeV}\, , ~~~ 10^{-4} \leq g_X \leq 1 \, . 
\end{equation}
The $g_X$ range is motivated by Eq.~(\ref{eq:gauge_masses}), requiring
the tree level $\rho$ parameter to be close to 1. The lower bound on
$t_\beta$ is due to the $b \to s$ constraints, while we choose the
upper bound (which corresponds to $v_1 = 1$ GeV) and the $v_3$ range
in such a way that we have $v_1 \ll v_3$ for most of the points. Using
the parameters in the ranges given in Eq.~(\ref{eq:range_parameters})
we get $10 {~\rm MeV}\lesssim m_X \lesssim 1$ TeV.

\begin{figure}[t!]
\centering
\setcounter{subfigure}{0}
\includegraphics[width=1\textwidth]{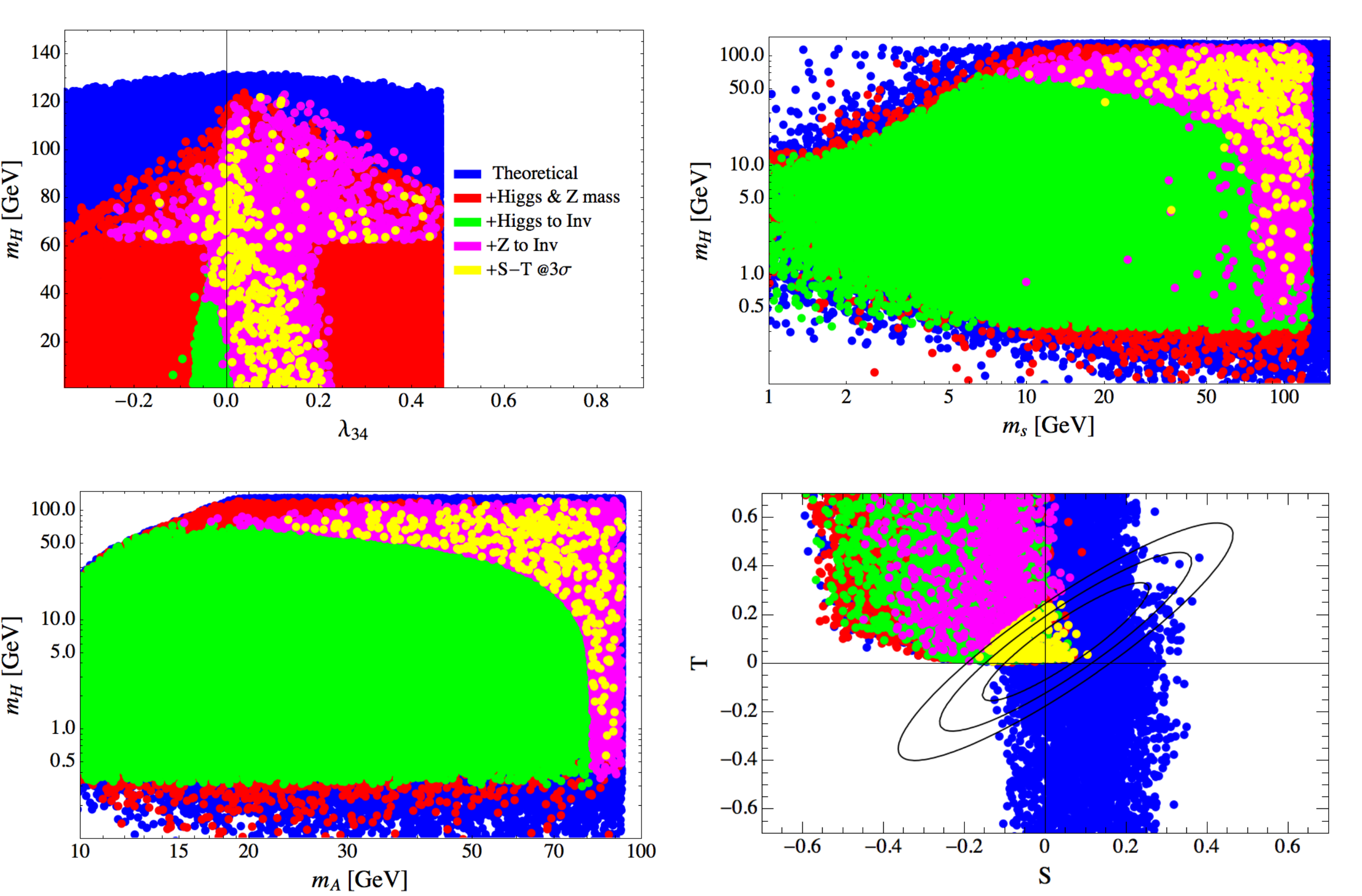}
 \label{fig1}
\caption{\label{fig:results} Effect of the theoretical and
  experimental constraints on the parameter space of the $\nu$2HDSM.
  We show in blue the points allowed by the theoretical constraints
  (Eqs.~(\ref{eq:perturbativity})-(\ref{eq:vac_stab})) and
  $m_{H^\pm}>$ 80 GeV, in red those that also give the correct values
  for $m_h$ and $m_Z$, in green points that in addition are allowed by
  $h \to \; \rm {invisible}$, in magenta those that also pass the
  limits on the $Z$ invisible width and finally in yellow those within
  them that are compatible with $S,T$ and $U$ at 3$\sigma$.  Top left:
  $(\lambda_{34}, m_H)$. Top right: $(m_s, m_H$).  Bottom left: $(m_A,
  m_H$). Bottom right: $(S, T)$.  }
\end{figure}

The results of the scan are presented in Fig.~\ref{fig:results}. The
color code is as follows: {\it (blue):} points that pass the
theoretical constraints of
Eqs.~(\ref{eq:perturbativity})-(\ref{eq:vac_stab}), as well as the
experimental limit $m_{H^\pm}>$ 80 GeV; {\it (red):} points that also
give values for $m_h$ and $m_Z$ within the limits described above;
{\it (green):} points that in addition are allowed by the limits on
the Higgs invisible decays (see Eqs.~(\ref{eq:htoSS})-(\ref{hAV}));
{\it (magenta):} points also allowed by the limits on the $Z$
invisible width (see Eq.~(\ref{GammaZ})); {\it (yellow)}: points
compatible with the EWPM limits of Eq.~(\ref{eq:STU}) at 3$\sigma$.

%%%%%%%%%%%%%%%%%%%%%%%%%%%%%%%%%%%%%%%%%%%%%%%%%%%%%
%%%%%%%%%%%%%%%%%%%%%%%%%%%%%%%%%%%%%%%%%%%%%%%%%%%%%
%%**************************************************   
%%%%%%%%%            figure 1         %%%%%%%%%%%%%%
%%**************************************************   
%%%%%%%%%%%%%%%%%%%%%%%%%%%%%%%%%%%%%%%%%%%%%%%%%%%%%
%%%%%%%%%%%%%%%%%%%%%%%%%%%%%%%%%%%%%%%%%%%%%%%%%%%%%

In the upper left panel we show the different points in the
$(\lambda_{34}, m_H)$ plane. The preference of the experimentally
allowed points for positive values of $\lambda_{34}$ is due to the
alignment limit of Eq.~(\ref{eq:alignment}), since $\lambda_{1,2}$ are
always positive due to the vacuum stability constraint (see
Eq.~(\ref{eq:vac_stab})). We can now clearly see that two regions
appear after imposing the bounds on $h \to$ invisible (and are
maintained by the subsequent bounds). The first one corresponds to
$m_A$, $m_H \lesssim m_h/2$, {\it i.e.} the region in which the
constraint of Eq. (\ref{eq:constr_lambda34}) applies. This forces
$\lambda_{34}$ to be below $0.25$. When the $h \to HH$ and $h \to AA$
decays close, larger values of $\lambda_{34}$ are allowed. In this
region, the upper bound on $m_H$ is due to the upper bound on $v_1$
coming from the Higgs mass in Eq.~(\ref{scalarMasses}).

Turning to the upper right and the lower left panels of
Fig.~\ref{fig:results}, we have that, in addition to the bounds on
$m_H$ already discussed, a lower and an upper bound on the $s$ and $A$
masses appear for the experimentally allowed points. The upper bound
on $m_s$ comes from the Higgs mass constraint, while the upper bound
on $m_A$ is due to the fact that it grows with $v_3$ (see
Eq.~(\ref{scalarMasses})) and is subject to the upper bound given in
Eq.~(\ref{eq:range_parameters}).  The lower bound (which depends 
on the value of $m_H$ and appears after the $Z \to$ invisible limit is
imposed) is instead due to the requirement $m_A + m_H > m_Z$ already
discussed.

Finally, we show in the right lower panel of Fig.~\ref{fig:results}
the points in the $(S,T)$ plane. The dominant contribution to $S$ and
$T$ is due to the charged scalar (whose mass, for our choice of
parameters and after applying all the bounds, is in the range $80 \, {\rm
  GeV} \lesssim m_{H^\pm} \lesssim 600\, {\rm GeV}$), and pushes $T$
to positive and $S$ to negative values. A surprising feature of the
allowed points is that, somewhat counterintuitively, a light $H$
scalar with mass $1$ GeV $\lesssim m_H \lesssim 20$ GeV does not give
a too large contribution to EWPM. This is due to the fact that we are
very close to the alignment limit, with the $H$ scalar almost
completely decoupled from the $Z$ and $W$ gauge bosons.

%%%%%%%%%%%%%%%%%%%%%%%%%%%
%%%%%%%%%%%%%%%%%%%%%%%%%%%
%%**************************************************   
%%%%%%%%%            figure 2         %%%%%%%%%%%%%%
%%**************************************************   
\begin{figure}[t!]
\centering
\setcounter{subfigure}{0}
\includegraphics[width=0.85\textwidth]{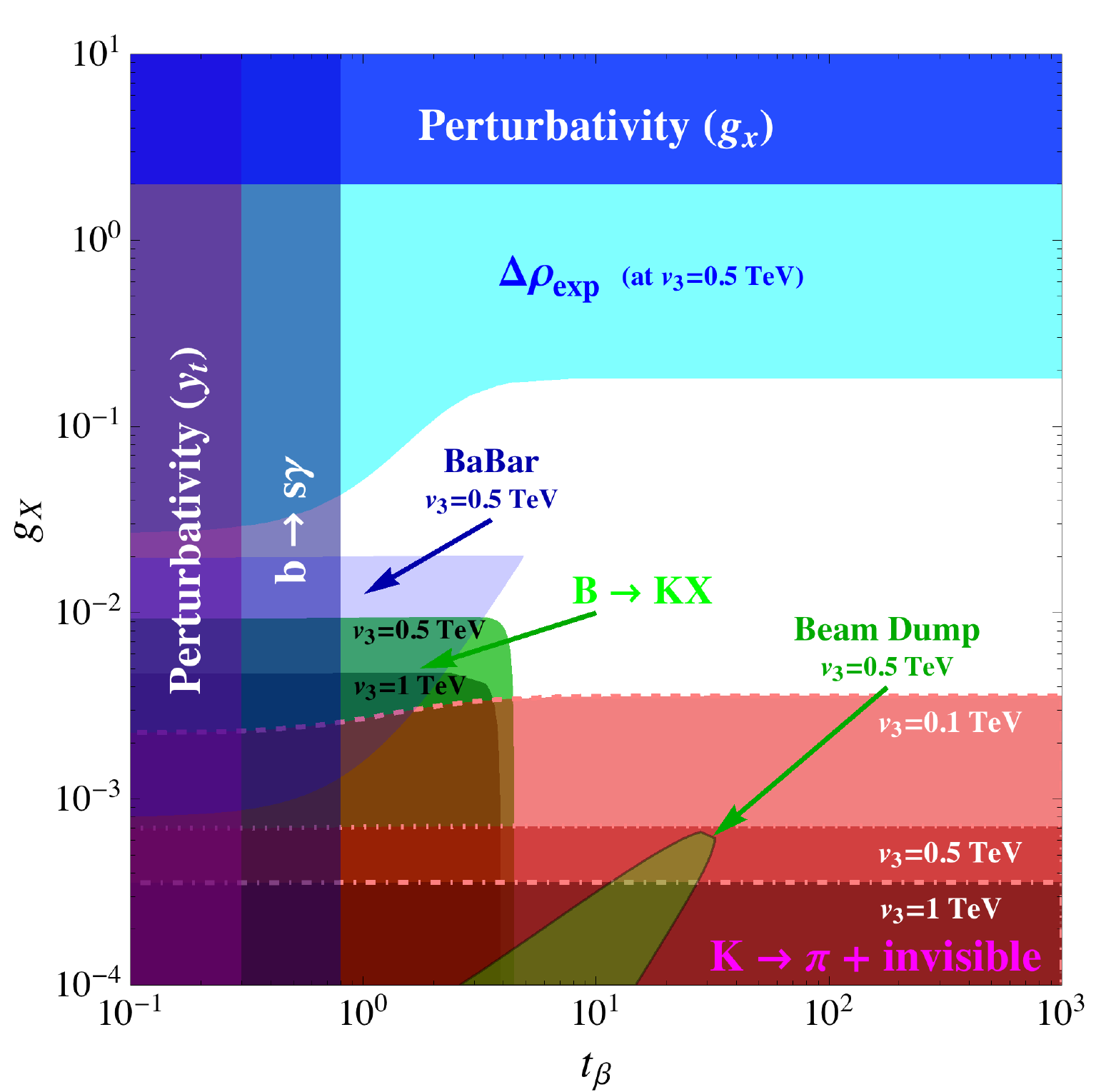}
 \label{fig1}
\caption{\label{fig:moneyplot}
Bounds on the $(t_\beta , g_X)$ plane coming from different theoretical and experimental 
sources. The colored regions are excluded.} 
\end{figure}
%%%%%%%%%%%%%%%%%%%%%%%%%%%%%%%%%%%%%%%%%%%%%%%%%%%%%
%%%%%%%%%%%%%%%%%%%%%%%%%%%%%%%%%%%%%%%%%%%%%%%%%%%%%

Let us conclude by showing, in Fig.~\ref{fig:moneyplot}, the bounds on
the $(t_\beta,g_X)$ plane. The lower bound on $t_\beta$ comes from the
upper bound on $m_{H^{\pm}}$ obtained from exclusive $b\to s\gamma$
decays which result in excluding the region $t_\beta \lesssim 0.8$ for
$m_{H^\pm}\lesssim600$~GeV~\cite{Misiak:2017bgg}. A milder lower bound
is provided instead by the theoretical requirement of a perturbative
top Yukawa coupling, eliminating $t_\beta \lesssim 0.3$. Turning to
$g_X$, we have that upper bounds can also be set by the stipulation of
having a perturbative $g_X$ at the scale $\Lambda$ and by asking for
compatibility with the observed value of $\rho$. In particular, the
cyan region is excluded by imposing $\rho-1=\hat{\alpha}(M_Z)T =
\frac{1}{127}(0.09 - 0.26)$~\cite{Olive:2016xmw}.  The lower bounds
are, however, more interesting.  As already explained in
Sec.~\ref{sec:constraints}, they come from the processes $K^\pm \to
\pi^\pm + {\; \rm invisible}$ and $B^\pm \to K^\pm + {\; \rm
  invisible}$ which, in our case, can be mediated by the $X$
boson. When a model allows for an extra neutral gauge boson with mass
mixing with the SM $Z$, these decay modes will give rise to severe
experimental constraints on this mixing. In our case, after applying
all the theoretical and experimental restrictions to our model, we get
an upper bound on the mass mixing parameter $\delta\lesssim0.07$ (see
Eq.~(\ref{delta})), a slightly larger value than the experimentally
allowed one.  Therefore, we need to close the aforementioned kaon and
$B$ decays and as a result, we get lower bounds on $m_X$ and $g_X$.  
As the excluded lower regions in Fig.~\ref{fig:moneyplot} show, these constraints
strongly depend on the assumed value of $v_3$. $B$ decays depend also on $m_{H^+}$ and  may 
impose a limit on $t_\beta$, on the other hand, kaon decays basically only control $g_X$.
As in our model we have
\begin{equation}
\delta\sim10^{-3}\frac{(v_1/100~{\rm{GeV}})^2}{v\sqrt{(v_1/100~{\rm{GeV}})^2+4\,(v_3/20~{\rm{TeV}})^2}}\, ,
\end{equation}
we see that for values of $v_1<100$~GeV, the restrictions from both meson decays 
are lifted if $v_3>20$~TeV and hence, in this case, $g_X$ can be as small as possible. 
Note that for such a large $v_3$ one would have to consider a much larger breaking scale $\Lambda$. 

Finally, since $X$ can couple to the SM fermions we also get limits on
$g_{Xf\bar{f}}\sim g_X c^2_\beta$ (see Eq. (\ref{gXff})). The main
constraint comes from beam-dump experiments (light green region)
which mainly affects low masses ($m_X\lesssim1$~GeV) and, consequently,
$g_X$ couplings in a range already excluded by
$K^\pm \to\pi^\pm + {\; \rm invisible}$. On the other hand, the BaBar
experiment, which is sensitive to $0.02~$GeV$<m_X<10.2$~GeV, excludes
$g_{Xf\bar{f}}>4\times10^{-4}$ \cite{Lees:2014xha} (light
blue region). Since these experiments are responsive to $m_X \lesssim 10$ GeV
the strongest constraints appear for smaller values of $v_3$.  The limit from 
$\Upsilon\to\gamma X$ decay, discussed in Sec.~\ref{sec:constraints}, is omitted here as it 
does not exclude any new region of the parameter space that is not already covered by other bounds.

%%%%%%%%%%%%%%%%%%%%%%%%%%%%
%%%%%%%%%%%%%%%%%%%%%%%%%%%%%
\section{Conclusions}\label{sec:conclusion}

The origin of neutrino masses is a long standing puzzle in particle
physics. In this work, we have proposed a new realization of the
inverse seesaw mechanism based on a {\it neutrinophilic}
Two-Higgs-Doublet Model, with the $U(1)_X$ symmetry introduced to
forbid a Majorana mass for the right-handed neutrinos promoted to a
gauge symmetry. As we have shown, the minimal particle content that
allows for a spontaneous breaking of the $U(1)_X$ and that cancels
anomalies is precisely what is needed to construct an inverse seesaw
model with the addition of a Majorana mass for the right-handed
neutrinos. We have then focused on the viability of the model,
analyzing the parameter space which is allowed by experiments. Our
main results are shown in Fig.~\ref{fig:results}, which show the
allowed values of the parameters in the scalar/gauge sector. As can be
seen, the parameter space is strongly constrained by the Higgs and $Z$
boson decays and by electroweak precision tests. Nevertheless, there
is still a considerably large region where the model is perfectly
viable. Some of this region may be scrutinized in the future with
improvements of the $h$ to invisible decay width measurements.  In
addition, $K^\pm \to \pi^\pm \, + $ invisible and $B^\pm \to K^\pm \,
+ $ invisible as well as dark photon searches in $e^+e^-$ and beam
dump experiments already place lower bounds on $g_X$, the new gauge
boson coupling.  Future experiments (like CERN
Na62~\cite{Ruggiero:2013oxa} and Fermilab
ORKA~\cite{Worcester:2012rd}) should be able to improve the
sensitivity to these meson decays by at least one order of magnitude,
testing values of $g_X$ that would otherwise only be accessible if a
future high energy lepton collider like the FCC-ee would be built.
The proposed SHiP experiment at CERN~\cite{Bonivento:2017acp} may also
be able to extend the beam dump limits to higher values of $m_X$.
These lower bounds on $g_X$, however, depend on the assumed value of
$v_3$.  In the neutrino sector we do not expect major changes with
respect to what happens for the usual implementation of the (double)
inverse seesaw mechanism. In contrast, what are the constraints that
might be obtained by supplementary requiring a successful leptgenesis
and/or the presence of a good dark matter candidate is not simple to
predict. This will be analyzed in a separate publication.

\acknowledgments This work was supported by Funda\c{c}\~ao de Amparo
\`a Pesquisa do Estado de S\~ao Paulo (FAPESP) and Conselho Nacional
de Ci\^encia e Tecnologia (CNPq). This project has also received
funding from the European Union's Horizon 2020 research and innovation
programme under the Marie Sklodowska-Curie grant agreement No 674896.
This manuscript has been authored by Fermi Research Alliance, LLC under Contract No. DE-AC02-07CH11359 with the U.S. Department of Energy, Office of Science, Office of High Energy Physics. The United States Government retains and the publisher, by accepting the article for publication, acknowledges that the United States Government retains a non-exclusive, paid-up, irrevocable, world-wide license to publish or reproduce the published form of this manuscript, or allow others to do so, for United States Government purposes.

\appendix
\section{Some detail on the scalar and gauge masses}\label{app:details}
%%%%%%%%%%%%%%%%%%%%%%%%%%%
%%%%%%%%%%%%%%%%%%%%%%%%%%%
In this appendix we present some useful detail about the
diagonalization of the scalar and mass matrices. Minimizing the
potential in Eq.~(\ref{potential}) we find that the mass matrices for
the CP-even and CP-odd scalars (defined as in Eq.~(\ref{doublets}))
are given by
\begin{eqnarray}\label{Mrho}
{\cal M}_{\rm CP-even}^2=
\begin{pmatrix}
\lambda_1v_1^2& \lambda_{34}v_1 v_2 +\frac{\sqrt{2}g_{\rm{NP}}v_1 v_2 v_3}{\Lambda}&\lambda_{1s}v_1 v_3+\frac{g_{\rm{NP}}v_1^2 v_2}{\sqrt{2}\Lambda}\\
  \lambda_{34} v_1v_2+\frac{\sqrt{2}g_{\rm{NP}}v_1 v_2 v_3}{\Lambda}&  \lambda_2v_2^2  &\lambda_{2s}v_2 v_3+\frac{g_{\rm{NP}}v_1 v_2^2}{\sqrt{2}\Lambda}\\
  \lambda_{1s}v_1v_3+\frac{g_{\rm{NP}}v_1^2 v_2}{\sqrt{2}\Lambda}&\lambda_{2s}v_2v_3+\frac{g_{\rm{NP}}v_1 v_2^2}{\sqrt{2}\Lambda}&-\frac{g_{\rm{NP}}v_1^2 v_2^2}{2\sqrt{2}\Lambda v_3}+\lambda_s v_3^2
\end{pmatrix} \, ,
\end{eqnarray}
%%
%%
%{\bf CP-even matrix}\\
and
\begin{eqnarray}\label{Meta}
{\cal M}_{\rm CP-odd}^2=\frac{g_{\rm{NP}}}{\Lambda}
\begin{pmatrix}
-\sqrt{2}v_1^2 v_3&\sqrt{2}v_1 v_2 v_3&-v_1^2 v_2/\sqrt{2}\\
\sqrt{2}v_1 v_2 v_3&-\sqrt{2}v_2^2  v_3&v_1 v_2^2/\sqrt{2}\\
-v_1^2 v_2/\sqrt{2}&v_1 v_2^2/\sqrt{2}&-\frac{v_1^2 v_2^2}{2\sqrt{2} v_3}
\end{pmatrix}.
\end{eqnarray}
The CP-odd matrix is easily diagonalized; the only massive state (the
{\em would-be Majoron} $A$) gets its mass via the explicit $U(1)_\ell$
breaking term, and we have
\begin{equation}
	\begin{array}{l}
		\ds{m_A^2} = \ds{-\frac{g_{NP}}{\Lambda} \frac{v_1^2 v_2^2 + 4 v^2 v_3^2}{2\sqrt{2} v_3^2}\, ,} \\
		\ds{A} = \ds{\frac{v_1 v_2}{\sqrt{v_1^2 v_2^2 + 4 v^2 v_3^2}} \left( \frac{2 v_3}{v_1} \eta_1 - \frac{2 v_3}{v_2} \eta_2 + \eta_3  \right)\, .}
	\end{array}
\end{equation}
In order to diagonalize the CP-even squared mass matrix we instead use perturbation theory.  
Defining the Higgs basis according to 
\begin{equation}\label{eq:Higgs_basis}
	h_{SM} = \frac{v_1}{v} \rho_1 + \frac{v_2}{v} \rho_2\, , ~~~~ H_{SM} = -\frac{v_2}{v} \rho_1 + \frac{v_1}{v} \rho_2\, ,
\end{equation}
and the mass eigenbasis as
\begin{equation}\label{eq:CPeven_true_rotation}
 \begin{pmatrix}
	 h_{SM}\\
	 H_{SM}\\
	 \rho_3
 \end{pmatrix} = U 
 \begin{pmatrix}
	 h \\
	 H \\
	 s
 \end{pmatrix}\, ,
\end{equation}
we get, to first order in the perturbation parameters ${\cal M}_{13, 23}$,
\begin{equation}\label{eq:CPeven_true_rotation2}
	\begin{array}{rcl}
		h & \simeq & c_\alpha h_{SM} + s_\alpha H_{SM} + U_{31} \rho_3 \, \\
		H & \simeq & -s_\alpha h_{SM} + c_\alpha H_{SM} + U_{32} \rho_3 \, \\
		s & \simeq & U_{13} h_{SM} + U_{23} H_{SM} + \rho_3 \, ,
	\end{array}
\end{equation}
where all the elements $U_{ij}$ with either $i=3$ or $j=3$ are ${\cal O}({\cal M}_{13, 23})$ and $c_\alpha$, $s_\alpha$ are the rotation elements computed diagonalizing the $2\times 2$ block involving $h_{SM}$ and $H_{SM}$. The masses computed with this approximation are presented in Eq.~(\ref{scalarMasses}). 

Let us now turn to the gauge sector. The covariant derivatives in the scalar sector are given by
\begin{eqnarray}\label{covariant}
D_{\mu}\Phi_1&=&(\partial_{\mu}-i\frac{g}{2}W^i_{\mu}\tau^i-i\frac{g^{\prime}}{2}B_{\mu}-i\frac{g_X}{2}X^0_{\mu})\Phi_1,\nonumber\\
D_{\mu}\Phi_2&=&(\partial_{\mu}-i\frac{g}{2}W^i_{\mu}\tau^i-i\frac{g^{\prime}}{2}B_{\mu})\Phi_2,\\
D_\mu S&=&(\partial_{\mu}-ig_X X^0_{\mu})S\nonumber,
\end{eqnarray}
%%
%
%
% For the neutral part of the gauge bosons, we have
%%
%%
The squared mass matrix of the neutral gauge bosons is 
\begin{eqnarray}\label{neutralGauge}
\mathcal{L}_{\rm{gauge}}^{\rm m}&=&
\frac{1}{8}
\begin{pmatrix}
B_{\mu}&W_{\mu}^3&X^0_{\mu}
\end{pmatrix}
\begin{pmatrix}
{g^{\prime}}^2(v_1^2+v_2^2)&-gg^{\prime}(v_1^2+v_2^2)&g^{\prime}g_X v_1^2\\
-gg^{\prime}(v_1^2+v_2^2)&g^2(v_1^2+v_2^2)&-gg_X v_1^2\\
g^{\prime}g_X v_1^2&-gg_X v_1^2&g^2_X(v_1^2+v_3^2)
\end{pmatrix}
\begin{pmatrix}
B_{\mu}\\
W_{\mu}^3\\
X^0_{\mu}
\end{pmatrix}\, ,
\end{eqnarray}
which can be diagonalized in two steps. First of all, we can define the standard photon field according to 
\begin{equation}\label{cw}
\begin{pmatrix}
W^3_{\mu}\\
B_\mu
\end{pmatrix}
=\begin{pmatrix}
c_W&s_W\\
-s_W&c_W
\end{pmatrix}
\begin{pmatrix}
Z^0_\mu\\
A_{\mu}
\end{pmatrix} \, .
\end{equation}
At this point, we can further rotate the $Z^0$ and $X^0$ fields to go to the mass eigenbasis:
\begin{equation}\label{ct}
\begin{pmatrix}
Z^0_\mu\\
X^0_{\mu}
\end{pmatrix}
=\begin{pmatrix}
c_t&s_t\\
-s_t&c_t
\end{pmatrix}
\begin{pmatrix}
Z_{\mu}\\
X_\mu
\end{pmatrix},
\end{equation}
where $c_t=\cos \theta_t$ and $s_t=\sin \theta_t$, so that 
\begin{equation}\label{tant}
\tan 2t=\frac{2g_X c_\beta^2\sqrt{g^2+{g^\prime}^2}}{g^2+{g^\prime}^2-g_X^2 c_{\beta^\prime}^{2}}.
\end{equation}
The masses of the neutral gauge bosons are given in
Eq.~(\ref{eq:gauge_masses}), with the first order expansion for $g_X
v_1 \ll g v$ given in Eq.~(\ref{gaugeMasses}).

\end{document}